\documentclass[aps,floats,showpacs]{revtex4}

\usepackage{graphicx}
 \usepackage{epsfig}
 \usepackage{pstricks}
 \usepackage{pst-coil}
\def\PRL#1{{ Phys.\ Rev.\ Lett.} {\bf #1}}
\def\PRD#1{{ Phys.\ Rev.} {\bf D#1}}
\def\NPB#1{{ Nucl.\ Phys.} {\bf B#1}}
\def\PLB#1{{Phys.\ Lett.} {\bf B#1}}
\def\vev#1{\langle #1 \rangle}
\def\be{\begin{equation}}
\def\ee{\end{equation}}
\def\bea{\begin{eqnarray}}
\def\eea{\end{eqnarray}}

\def\sss{\scriptscriptstyle}
\def\Ls{{\sss L}}
\def\Ms{{\sss M}}
\def\Rs{{\sss R}}
\def\Ns{{\sss N}}

\def\Tr{\rm Tr}
\def\dld{\Delta_\Ls^\dagger}
\def\dl{\Delta_\Ls}
\def\drd{\Delta_\Rs^\dagger}
\def\dr{\Delta_\Rs}
\def\p{\phi}
\def\pd{\phi^\dagger}
\def\pti{\tilde\phi}
\def\ptid{\pti^\dagger}
\def\Bp{B_{\sss +}}
\def\Bm{B_{\sss -}}
\def\Ap{A_{\sss +}}
\def\Am{A_{\sss -}}
\def\lm{\lambda_{\sss -}}
\def\lp{\lambda_{\sss +}}
\def\rhm{\rho_{\sss -}}
\def\rhp{\rho_{\sss +}}
\def\kbd{\left({m\over M}\right)^2}
{\newcommand{\lsim}{\mbox{\raisebox{-.6ex}{~$\stackrel{<}{\sim}$~}}}
{\newcommand{\gsim}{\mbox{\raisebox{-.6ex}{~$\stackrel{>}{\sim}$~}}}
\def\DT{\Delta_T}
\def\GD{\Gamma_{\sss D}}
\def\GS{\Gamma_{\sss S}}

\begin{document}

\title{Transient domain walls and lepton asymmetry in the Left-Right 
symmetric     model}

\author{J.\ M.\  Cline}
\affiliation{McGill University, 3600 University St.
Montr\'eal, Qu\'ebec H3A 2T8, Canada \\
E-mail: jcline@physics.mcgill.ca}

\author{U.\  A.\  Yajnik,  S.\ N.\ Nayak\footnote{
address after September 2001 : Physics Department, Sambalpur University,
Sambalpur 768017, Orissa}
 and M.\ Rabikumar}
\affiliation{Indian Institute of Technology Bombay, Mumbai
400\thinspace076,
India \\
E-mail: yajnik@phy.iitb.ac.in, rabi@phy.iitb.ac.in}

\date{April 25 2002}

\begin{abstract}
It is shown that the dynamics of domain walls in Left-Right 
symmetric models, separating respective regions of unbroken 
$SU(2)_L$ and $SU(2)_R$ in the early universe,  can give rise to
baryogenesis via leptogenesis. Neutrinos have a spatially varying 
complex mass matrix due to CP-violating scalar condensates 
in the domain wall.  The motion of the wall through the plasma 
generates a flux of lepton number across the wall which is
converted to a lepton asymmetry by helicity-flipping scatterings.
Subsequent processing of the lepton excess by sphalerons results
in the observed baryon asymmetry, for a range of parameters in 
Left-Right symmetric models.
\end{abstract}


\maketitle

\section{Introduction}\label{sec:intro}

Explaining the observed baryon asymmetry of the Universe within the
framework of gauge theories and the standard Big Bang cosmology remains
an open problem. The study has resulted in a  deeper understanding of
nonperturbative phenomena at finite temperature in gauge theories
including supersymmetric theories. Many of the particle physics models and
scenarios considered so far seem to require unnatural extensions or
very special choices of parameters for successful baryogenesis; prime
among these are the standard model (SM) and its  minimal supersymmetric
extension (MSSM), using a first order phase transition  \cite{shap,CKN,
others}.  Among the alternative proposals are those which rely on
the presence of topological defects, viz., domain walls \cite{mencoo}, 
and cosmic strings \cite{sbdanduay1,branden1}. 
The latter are generic to many gauge theories. What makes them
especially suited for baryogenesis is their nonthermal nature soon
after their formation. Unlike the need for a first order phase
transition which sets severe limitations on the couplings and
particle content of the model, existence of defects relies only on
topological features of the  vacuum manifolds and permits nonthermal
effects without fine tuning.

Many special features arise when studying cosmological consequences of
topological defects in any given gauge model. The Left-Right symmetric
(L-R) model was studied in the context of conventional baryogenesis 
mechanisms in \cite{mohzha, Frere:1992bd} and in the context of 
domain wall mediated mechanism in \cite{lewrio}. A detailed study of 
the possible defects existing in the L-R model 
was made in \cite{ywmmc}. It was argued that the
domain wall configurations implied by the symmetry breaking pattern
present possibilities for baryogenesis. In this paper we study the
interaction of neutrinos which derive Majorana masses from the scalar
condensate which constitutes the domain wall. Many of the broad
features encountered, {\it e.g.,} asymmetric reflection and
transmission of fermions from moving domain walls,  have appeared in
the study of electroweak baryogenesis.
In the diffusion-enhanced scenario \cite{CKN} driven by thick
walls, the asymmetry diffusing in front of the wall is equlibrated 
by high temperature  sphalerons. In our mechanism this is replaced by
helicity flipping interactions in front of the wall which arise from
the scalar condensate imparting a Majorana mass to the fermions.
Our parametric answer for the unprocessed Lepton asymmetry produced in this
mechanism is therefore dominated by the Majorana mass parameter
$f$ in eq. (\ref{eq:ansparam}).
The scalar condensate is absent behind the wall and therefore the
asymmetry that has streamed through persists. 

At the completion of the $L-R$ symmetry breaking transition, a 
particular hypercharge ${\tilde Y}=I^3_R-\frac{1}{2}(B-L)$ 
is demonstrated to be spontaneously generated in the form 
of left-handed neutrinos.  Due to the high-temperature
electroweak sphalerons, which set $B+L=0$, this will be converted
into an asymmetry of baryon minus lepton number ($B-L$). 
The baryon asymmetry thus generated arises in addition to that from the
well-known leptogenesis  \cite{fukyan,luty} mechanism  due to
Majorana neutrinos. However, the two mechanisms constrain the 
Left-Right model rather differently. The usual mechanism requires the 
$Z_R$ mass to be larger than the heavy neutrino mass  \cite{luty,
plum,buchplum}. The present mechanism constrains the parameters 
of the Higgs sector for adequate $CP$ violation, and the Majorana
Yukawa couplings as already pointed out.  Our main result is the 
identification of broad ranges of these parameters that ensure
sufficient lepton asymmetry. Further, the subsequent evolution of 
this asymmetry must successfully produce the abserved baryon asymmetry.
This requirement can be used to constrain the temperature scale of the
$L-R$ phase transition, eq. (\ref{limit1}), or alternatively, the 
light neutrino  mass, eq. (\ref{limit2}).

In \cite{Harvey:1990qw} and \cite{Fischler:1990gn} the possibility 
of $B-L$ generated by any mechanism being neutralised due to 
presence of heavy majorana neutrino was considered. 
The bound first obtained in \cite{Harvey:1990qw} is
$20M_{\sss N}\ \gsim\  \sqrt{T_{B-L}M_{Pl}}$ with $M_{\sss N}$
the mass scale of the heavy majorana neutrino and $T_{B-L}$ the
temperature at which $B-L$  originates. This is derived from
lepton number depletion due to heavy neutrino mediated scattering
processes and assumes $T_{B-L} >M_{\sss N}$. 
It was argued in \cite{Fischler:1990gn} that 
the requirement that the heavy neutrino decays occur only in
out of equilibrium conditions places a more stringent bound.
Using the see-saw relation, it requires 
\be
m_{\nu}\lsim m_*\equiv 4\pi g_*^{1/2}{G_N^{1/2} \over \sqrt2 G_F} 
\sim 10^{-3} eV.
\ee
where $G_N$ is the Newton constant and $G_F$ is the Fermi constant. 
Current neutrino data easily suggest a larger neutrino mass.
In this case it is argued that \cite{Fischler:1990gn}  
one needs 
\be 
T_{B-L}\lsim M_{\sss N}= {h^2\over 2}\left( m_*\over m_\nu \right)^2 
10^{17} GeV
\ee
These
considerations generically need a low scale for $B-L$ creation.
Detailed investigations\cite{Buchmuller:2000as}
of Leptogenesis scenrios, including
lepton generation mixing, show that in several specific unified models 
this can be achieved in the context of conventional Leptogenesis. 
The present mechanism has the potential of meeting the requirement 
of low scale $B-L$  generation in a natural way, although 
detailed investigations remain to be carried out.
We shall return to this point in sec. \ref{sec:cosmology}.

The paper is organised as follows. Section \ref{sec:model}  introduces 
the main features of the Left-Right symmetric model, synthesizing the
conventions used by previous authors with the ones we follow. Section
\ref{sec:mechanism} discusses the microscopic mechanism of lepton 
number violation in
scattering near the domain wall.  It outlines the method that can be
used for detailed study of lepton number creation in this context.
Section \ref{sec:walls} demonstrates the existence of the conditions 
required for
lepton number creation, in particular the CP-violating nature of the
wall profiles. Section \ref{sec:toymodel} presents a  simplified 
version of the full
theory to be studied and numerical results justifying the general
conclusions of the previous section. Section \ref{sec:cosmology}
 discusses the
implications to cosmology. Overall conclusions are presented in the
last section.

\section{The Left-Right symmetric model}\label{sec:model}

For the purpose of model building, Left-Right symmetry is a broad 
category, with several possible implementations. 
In this paper we shall adopt its more
popular version which is desribed below. From the point of view
of our mechanism, the discrete symmetry under exchange of the
$SU(2)_R$ field content of the model with $SU(2)_L$ field content 
is crucial. The breakdown of this symmetry gives rise to domain 
walls whose field configuration we study in detail. The most elegant
version of the model consists of identical values of the
two $SU(2)$ gauge couplings in addition to a strict equality
of certain scalar couplings in the Higgs potential. This may seem like 
an artificial requirement, considering that the two semisimple groups 
are independent, and there are no dynamical hints why they must
be exactly same to begin with. More importantly, if this requirement
is imposed an unpleasant feature arises in the context of cosmology.
Breakdown of an exact discrete symmetry gives rise to stable
domain walls and unless some mechanism removes them, they quickly
come to dominate the energy density of the Universe, contrary to
observations. Thus departure from exact symmetry is in any case
a phenomenologically desirable feature.
Happily, the mechanism being proposed here works well so long as the
departure from exact symmetry is small so that 
domain walls indeed form as transient constructs. A quantitiative 
discussion of this is taken up in sec. \ref{sec:LRbrpt}.

We now recapitulate the minimal $SU(2)_L \otimes SU(2)_R \otimes U(1)_{B-L}$
model  \cite{Senjanovic2,mohapbook}.  Parity is restored above
an energy scale $v_R$,  taken to be much higher than the electroweak scale,
by introducing the $SU(2)_R$ gauge symmetry which breaks at $v_R$.
Accordingly, a right-handed heavy neutrino species is added to each 
generation, and the gauge  bosons consist of two triplets $W_L^\mu \equiv
(3,1,0)$, $W_R^\mu \equiv (1,3,0)$ and a singlet $B^\mu \equiv (1,1,0)$. A
Left-Right symmetric assignment of gauge SU(2) charges to the fermions shows
that the new hypercharge needed to obtain the usual electric charge correctly
is exactly $B-L$. It is appealing that in this model the weak hypercharge is
related to known conserved charges. 

The electric charge formula now assumes a
Left-Right symmetric form
\be
Q=T^3_{L} + T^3_{R} + \frac{B-L}{2} ~~ ,
\ee
where $T^3_L$ and $T^3_{R}$ are the weak isospin represented by
$\tau^3/2$, and $\tau^3$ is the Pauli matrix.

The Higgs sector of the model is dictated by two considerations: the
pattern of symmetry breaking and the small masses of the
known neutrinos via the seesaw mechanism. The minimal set to
achieve these goals is
\begin{eqnarray}
\phi = \pmatrix{\phi_1^0 & \phi_1^+ \cr \phi_2^- & \phi_2^0} \equiv
\pmatrix{\frac{1}{2},\frac{1}{2},0} ~~, \nonumber \\
\dl  = \pmatrix{\frac{\delta_\Ls^+}{\sqrt{2}} & \delta_\Ls^{\sss ++} \cr
\delta_\Ls^0 & -\frac{\delta_\Ls^+}{\sqrt{2}}} \equiv (1,0,2) ~~ ,
\nonumber \\
\dr = \pmatrix{\frac{\delta_\Rs^+}{\sqrt{2}} & \delta_\Rs^{\sss ++} \cr
\delta_\Rs^0 & -\frac{\delta_\Rs^+}{\sqrt{2}}} \equiv (0,1,2) ~~.
 \label{Higgs}
\end{eqnarray}
where the electric charge assignment of the component fields has been
displayed and the representation with respect to the gauge group
is given in standard notation.

The minimal form of the Higgs potential needed to fulfill the main
phenomenological requirements can be found in  \cite{mohapbook}. This is
however not the most general form.  In  \cite{dgko} as well as 
 \cite{barenber} the possibility of spontaneous CP violation was
considered. In this case the couplings are chosen to be real, yet the
translation invariant minimum of the potential occurs for complex VEV's
(vacuum expectation values). The absence of explicit CP-violating
couplings makes it easier to accommodate phenomenological constraints
on CP violation.  
In the cosmological context in which we treat this
theory, this motivation is not as compelling. Nevertheless, the same
simplifying assumption will be made here.

Consider the potential parameterized as
\footnote{Our conventions parallel those of  \cite{barenber} but our
$\beta_2$ and $\beta_3$ have been assumed equal in that reference.}
\be
V = V_\Phi + V_\Delta + V_{\Phi \Delta} ~~ , \label{potential}
\ee
with
\bea
V_\Phi  =  &-&   \mu_1^2 \Tr \p\pd  - \mu_2^2(\Tr \pti\pd
+ \Tr \p\ptid) \nonumber \\ [0.2cm]
&  + & \lambda_1(\Tr\p\pd)^2
+\lambda_2\{ (\Tr\pti\pd)^2 + (\Tr\ptid\p)^2 \} \nonumber \\ [0.2 cm]
& + &  \lambda_3(\Tr\pti\pd)(\Tr\ptid\p)
+\lambda_4\{\Tr\pd\p(\Tr\pd\pti + \Tr\ptid\p)\}
~~ ,
\label{vphi}
\eea
\bea
V_\Delta  = & - & \mu_3^2 \Tr\Big(\dld \dl + \drd \dr\Big)
 +
\rho_1 \Big[\Big(\Tr(\dld \dl)\Big)^2 + \Big(\Tr(\drd \dr)\Big)^2\Big]
 \nonumber \\ [0.3cm]
& + &
\rho_2 \Big[\Tr\dld\dld \Tr \dl\dl + \Tr\drd\drd \Tr\dr\dr \Big]
+
\rho_3 \Big[\Tr\Big(\dld \dl\Big) \Tr \Big(\drd \dr\Big)\Big] \nonumber
\\ [0.3cm]
& + &
\rho_4 \Big[\Tr\dl\dl\Tr\drd\drd
 +
\Tr\dld\dld \Tr\dr\dr\Big]   ~~ , \label{vdelta}
\eea
\bea
V_{\Phi\Delta} & = & \alpha_1\Tr\p\pd( \Tr\dl\dld + \Tr\dr\drd)
\nonumber \\ [0.2cm]
&+& \alpha_2 \left\{ \Tr(\ptid\p)\Tr(\dr\drd)
+ \Tr(\pti\pd)\Tr(\dl\dld) \right\} 
\nonumber \\ [0.2cm]
&+& \alpha_2^* \left\{ \Tr(\pti\pd)\Tr(\dr\drd)
+ \Tr(\ptid\p)\Tr(\dl\dld) \right\}
\nonumber \\[0.2cm]
& + & \alpha_3\left\{\Tr(\p\pd\dl\dld) + \Tr(\pd\p\dr\drd) \right\}
\nonumber \\[0.2cm]
& + &  \beta_1 \left\{\Tr(\p\dr\pd\dld) + \Tr(\pd\dl\p\drd) \right\}
+ \beta_2 \left\{\Tr(\pti\dr\pd\dld) + \Tr(\ptid\dl\p\drd)\right\}
\nonumber \\[0.2cm]
&+& \beta_3 \left\{\Tr(\p\dr\ptid\dld) + \Tr(\pd\dl\pti\drd)\right\}
 ~~ .
\label{vdeltaphi}
\eea
All the parameters except $\alpha_2$ in the above are required
to be real by imposing the discrete symmetry
\be
\dl \leftrightarrow \dr, \qquad \p \leftrightarrow \pd,
\label{lrexch}
\ee
simultaneously with the exchange of left-handed and right-handed
fermions. Finally, $\alpha_2$ is chosen to be real from the requirement
of spontaneous $CP$ violation \cite{dgko, barenber}.

The ansatz for the VEV's of the scalar fields has been
discussed extensively in the literature. 
After accounting for phases
that can be eliminated by global symmetries and field redefinitions
 \cite{dgko}, only two independent phases remain. We choose them
for convenience as follows in the translation invariant VEV's:
\be
\p = \pmatrix{k_1e^{i\alpha} & 0 \cr  0 & k_2}, ~~
\Delta_{\Ls} = \pmatrix{ 0 & 0 \cr v_{\Ls}e^{i\theta} & 0}, ~~
\Delta_{\Rs} = \pmatrix{ 0 & 0 \cr v_{\Rs} & 0} ~~,
\label{vevs}
\ee
where all the other parameters are taken to be real.
Phenomenologically the hierarchy $v_\Ls  \ll k_1, k_2$, $\ll v_\Rs$
is required. This separates the electroweak scale from the
L-R symmetry breaking scale. It has been argued  \cite{barenber} 
that this is possible
to achieve for natural values of the above parameter set
while also obtaining (1) spontaneous CP violation, (2) mixing
of $W_\Rs, Z_\Rs$ with their $SU(2)_L$ counterparts which is 
unobservable
at accessible energies, and
(3) suppression of flavor changing neutral currents.

Fermion masses are obtained from Yukawa couplings of  quarks and
leptons with the Higgs bosons. For one generation of quarks $q$ and
leptons $\psi$, the couplings are given by  \cite{mohapbook}
\begin{eqnarray}
{\mathcal L}_Y & =  & h^q \, \bar q_L \,  \phi \, q_R + \tilde h^q \,
\bar q_L \, \tilde\phi \, q_R   \nonumber \\
 & + & h^l \, \bar\psi_L\,  \phi \, \psi_R + \tilde h^l \, \bar\psi_L
 \, \tilde\phi \,  \psi_R  \nonumber \\
& + & f \left( \psi_L^T \, C^{-1} \, \tau_2 \, \Delta_L \, \psi_L +
 \psi_R^T \,  C^{-1} \, \tau_2 \, \Delta_R \, \psi_R \right) +
 {\rm h.c.} \label{Yukawa}
\end{eqnarray}
where $C$ is the charge-conjugation matrix ($\psi^c = C\bar\psi^T$).
Neutrino mass terms resulting from the above parameterization of 
the scalar VEV's are 
\be
{\mathcal L}_{\nu-{\rm mass}} = \bar \nu_\Ls \, (h^lk_1e^{i\alpha}
+ \tilde h^lk_2)\nu_\Rs 
+ \left\{f\nu_{\Ls}^T\sigma^2 v_{\Ls}e^{i\theta}\nu_{\Ls}
+ f\nu_{\Rs}^T\sigma^2 v_{\Rs}\nu_{\Rs} + h.c. \right\}
\ee
The Majorana mass terms allowed for the neutrinos are a source of
lepton number violation, while  the CP violation needed for
leptogenesis results from the
phase $\alpha$ in the Dirac mass term.

\section{Leptogenesis mechanism}
\label{sec:mechanism}

Sources of CP violation as well as existence of out-of-equilibrium
conditions have been major challenges for realizing low energy
baryogenesis. The presence of moving topological defects in unified
theories is a novel source of out-of-equilibrium conditions. In ref.\
 \cite{ywmmc} it was shown that in the L-R model, at the first stage of
gauge symmetry breaking, domain walls can form,  which separate phases
of broken $SU(2)_R$ and $SU(2)_L$. The disappearance of the unstable
domains with unbroken $SU(2)_R$ provides a preferred direction for the
motion of the domain walls.  This can fulfill the
out-of-thermal-equilibrium requirement for leptogenesis.

Consider the interaction of neutrinos from the L-R wall, which is
encroaching on the energetically disfavored phase.  The left-handed
neutrinos, $\nu_\Ls$,  are massive in this domain, whereas they are
massless in the phase behind the wall. This can be seen from the
Majorana mass term $h_\Ms\Delta_\Ls{\overline{\nu_\Ls^c}}\nu_\Ls$, and
the fact that $\vev{\Delta_\Ls}$ has a kink-like profile, being zero
behind the wall and $O(v_\Rs)$ in front of it. 

To get leptogenesis, one needs an asymmetry in the reflection and
transmission coefficients from the wall between $\nu_\Ls$ and its CP
conjugate $(\nu_\Ls^c)$.  This can happen if a CP-violating condensate
exists in the wall. This comes from the Dirac mass terms as discussed 
in section \ref{sec:wallprofiles}. 
Then there will be a preference
for transmission of, say, $\nu_\Ls$. The corresponding excess of
antineutrinos $(\nu_\Ls^c)$ reflected in front of the wall will quickly
equilibrate with $\nu_\Ls$ due to helicity-flipping scatterings, whose
amplitude is proportional to the large Majorana mass.  However the
transmitted excess of $\nu_\Ls$ survives because it is not coupled to
its CP conjugate in the region behind the wall, where $\vev
{\Delta_L}=0$.

A quantitative analysis of this effect can be made either in the
framework of quantum mechanical reflection, valid for domain walls
which are narrow compared to the particles' thermal de Broglie
wavelengths, or using the classical force method  \cite{jpt,
clijokai,clikai}, which is gives the dominant contribution for walls
with larger widths. We adopt the latter here. The thickness
of the wall depends on the shape of the effective quartic potential
and we shall here treat the case of thick walls. Further, 
we assume that the potential energy difference between 
the two kinds of vacua is small, for example suppressed by Planck 
scale effects. In this case the pressure difference across the 
phase boundary is expected to be small, leading to slowly moving walls. 

In refs.\  \cite{jpt,
clijokai,clikai}, it is shown that the classical CP-violating force of 
the condensate on a fermion (in our case a neutrino) with
momentum component $p_x$ perpendicular to the wall is
\be
\label{eq:force}
F = \pm {\rm sign}(p_x)\frac{1}{2E^2}\left(m_\nu^2(x) 
\alpha'(x)\right)'.
\ee
The sign depends on whether the particle is $\nu_\Ls$ or $\nu_\Ls^c$, 
$m_\nu^2(x)$ is the position-dependent mass, \(E\) the 
energy  and $\alpha$ is
the spatially varying CP-violating phase. One can then derive
a diffusion equation for the chemical potential $\mu_\Ls$ of the
$\nu_\Ls$ as seen in the wall rest frame:
\be
\label{eq:diffeq}
-D_\nu \mu_\Ls'' - v_w \mu_\Ls'
+ \theta(x)\, \Gamma_{\rm hf}\,\mu_\Ls = S(x).
\ee
Here $D_\nu$ is the neutrino diffusion coefficient,
$v_w$ is the velocity of the wall, taken to be moving in the $+x$
direction, $\Gamma_{\rm hf}$ is the rate of helicity
flipping interactions taking place in front of the wall (hence
the step function $\theta(x)$), and $S$ is the source term,
given by
\be
\label{eq:source}
   S(x) = - {v_w D_\nu \over \vev{\vec v^{\,2}}} 
	\vev{v_x F(x)}',
\ee
where $\vec v$ is the neutrino velocity and
the angular brackets indicate thermal averages.
The net lepton number excess can then be calculated from
the chemical potential resulting as the solution of 
eq.\ (\ref{eq:diffeq}).

In order to use this formalism it is necessary to establish
the presence of a position-dependent phase $\alpha$. This
is what we turn to in the following discussion of the nature 
of domain walls in the L-R model.

\section{ Domain walls}\label{sec:walls}

\subsection{The Left-Right breaking phase transition}
\label{sec:LRbrpt}

The fundamental L-R symmetry of the model, eq.\ (\ref{lrexch}), 
implies that the gauge forces visible at low energies might have
been the $SU(2)_R$ rather than the $SU(2)_L$ with corresponding
different hypercharge remnant of the $U(1)_{B-L}$. In the early
Universe when the symmetry breaking is first signaled,
either $\dl$ or $\dr$ could acquire a VEV.
In mutually uncorrelated horizon volumes, this choice is 
random. As such we expect a domain structure with either of
these fields possessing a VEV in each domain. These may be
referred to as {\it $L$-like} if they lead to observed phenomenology
(with V-A currents),
and  {\it $R$-like}, if $\dr$ has remained zero.
Such domains will be separated by domain walls, dubbed
L-R walls in  \cite{ywmmc}.

The walls must disappear; otherwise they would contradict
standard cosmology by dominating the energy density very soon
after their formation.
This must occur in such a way as to eliminate the $R$-like regions.
What biases the survival of the $L$-like regions cannot 
be predicted within the model. We will assume that there are
small corrections suppressed by a grand unification scale mass, which
favour the $L$-like regions slightly. A time
asymmetry, due to the motion of the walls into the $R$-like regions,
arises as a result. The L-R walls necessarily
convert the $R$-like regions into $L$-like ones and disappear
by mutual collisions.

This can get implemented in two ways. One is explicit deviations 
from exact symmetry in the tree level Higgs lagrangian. An alternative
is that the gauge couplings of the two $SU(2)$'s are not identical. 
In this case the thermal perturbative corrections to the Higgs field free
energy will not be symmetric and the domain walls will be unstable.

A possible reason for such small deviations from exact
discrete symmetry could be that the model is actually 
descended from another unified model, and the small departures 
from exact symmetry are due to terms suppressed by 
the ratio of L-R breaking scale to the scale of higher 
unification. 
If the higher unification is in a conventional gauge group like $SO(10)$, 
it may not constitute a good explanation since the breaking of such 
symmetry  groups does not generically result in a low energy model 
with close-to-exact L-R symmetry.  It is however possible that
the unification is of a different type, for instance supergravity
or string unification, wherein mechanisms as yet not understood
impose the kind of symmetry required, while permitting small
energy differences in the free energies of the $L$-like versus 
$R$-like phases. A study of disappearance of domain walls in the 
context of a Supersymmetric model has been made in \cite{Abel:1995wk} 
and a study of the effectiveness of the mechancism in \cite{Larsson:1996sp}.

The breakdown of the L-R symmetry is  described by 
the VEVs of two scalars  $\dl$, $\dr$.  The form of the potential 
(\ref{vphi})-(\ref{vdeltaphi}) has been shown to have generic
zero temperature vacua which are either $R$-like or $L$-like. 
Let the difference in vacuum energy densities 
due to departure from exact $L$-$R$ symmetry be
$\delta U_{\Ls-\Rs}$ such that $L$-like vacuum is favoured.
If this difference is purely in the scalar self couplings,
it is determined directly by the GUT scale mechanism and will not 
be altered at finite temperature. On the other hand if the 
gauge couplings differ due to these suppressed GUT effects, the  
corresponding  thermal corrections will thereby acquire differences,
producing a corrected $\delta U^T_{\Ls-\Rs}$ at finite temperature. 
The condition for the
formation of the unstable domains can now be obtained as follows.
If the phase transition is  second order, its dynamics may be considered
to have terminated after the Ginzburg temperature $T_G$ is reached, which is
given by \cite{Kibble:1980mv}
\be
{(T_c-T_G)\over T_c} \simeq \lambda
\ee
where $T_c$ is the critical temperature and $\lambda$ the effective
quartic coupling. The correlation length  at this temperature
is estimated to be $\xi_G\simeq 1/(\lambda T_c)$.
For the walls to form, the fluctuations that can convert the
false vacuum to the true one must be suppressed
before the Ginzburg temperature is reached. Thus the energy excess
available within a correlation volume must be substantially less 
than the  energy needed to overcome the barrier set by $T_c-T_G$, {\it i.e.,}
\be
\delta U^T_{\Ls-\Rs}\,\xi_G^3\ \ll\ T_c-T_G \nonumber
\ee
or
\be
\delta U^T_{\Ls-\Rs}\ \ll\ \lambda^4 T_c^4\ \simeq\ \lambda^2 {V_T}^4  
\ee
where we took the temperture-dependent VEV
$V_T$ to be $\sim\lambda^{1/2}T_c$. This bound
 is easily satisfied if the GUT scale is much higher 
than the $L-R$ scale, as is expected.

\subsection{Wall profiles and CP violating condensate}
\label{sec:wallprofiles}

In order for nontrivial effects to be mediated by the
walls, the fermion species of interest should get a 
space-dependent mass from the wall. Furthermore, the CP-violating
phase $\alpha$ should also possess a nonvanishing gradient in the
wall interior. We study the minimization of the total
energy functional of the scalar sector with this in mind.

The minimization conditions for the various VEV's
introduced above are given in Appendix A, assuming
translational invariance. The presence of walls breaks
this invariance, requiring derivative terms to be added
in the minimization conditions.

We demonstrate that there are sizeable domains in the parameter
space for which a position-dependent, CP-violating condensate results.
In order to simplify the analysis we assume $k_1=k_2\equiv k$.
The range of the parameter values for which such minima would be
phenomenologically viable have been studied, {\it e.g.,}  in \cite{barenber}.
The analysis can be repeated for other cases along similar lines.
Let the L-R wall be located in the $y$-$z$ plane at $x=0$.
Its equation of motion is
\bea
\ddot{k} - k'' +  \left({d\alpha\over dx}\right)^2k &+&
 (-\mu_1^2 - 2\mu_2^2 \cos\alpha + \Delta\mu^2_{\sss T}) k \nonumber \\
&+& ( \alpha_2 + \frac{1}{4}\alpha_3 + \frac{1}{2}\alpha_1)
(v_\Ls^2 + v_\Rs^2)k  \nonumber \\
&+& \frac{1}{2}\{\beta_1\cos(\alpha-\theta) + \beta_3\cos\theta
+ \beta_2\cos(2\alpha-\theta)\}  v_\Ls v_\Rs k  \nonumber \\
&+& \{\lambda_1 + \lambda_2 (1+4\cos2\alpha) + \lambda_3 +
 4\lambda_4\cos\alpha\}  k^3 = 0 \ .
\label{keqn}
\eea
The temperature correction to the mass-squared term 
($\Delta\mu^2_{\sss T}$) is displayed
explicitly. The remaining parameters are also mildly 
temperature-dependent but this is a small effect.
The background fields $v_\Ls$, $v_\Rs$ have solutions of the
form  $\DT(1\pm \tanh(x/\Delta_w))$, with  upper and lower signs 
being for L and R respectively. 
$\DT$ is the temperature-dependent VEV which is possible for
either of $L$ or $R$ fields. This value is of the order of the temperature
$T$ relevant to the epoch immediately after the L-R breaking phase 
transition. $\Delta_w$ is the wall width, of the order 
$\DT^{-1}\lambda^{-1/2}$, $\lambda$ here standing for
the generic quartic coupling in the effective hamiltonian for the
$v_\Ls$ and  $v_\Rs$ fields. The nonderivative terms of this equation 
can be schematically written as
\be
m^2k + A(v_\Ls^2 + v_\Rs^2)k + Bv_\Ls v_\Rs k + Lk^3\ =\ 0 \ .
\label{schematic}
\ee
We are assuming $m^2>0$ so that at the epoch in question, $k=0$
in the absence of walls. This potential has two minima,
\be
k=0 \hskip 1cm {\rm or} \hskip 1cm k=k_{(0)}\equiv -{1\over L}\left( m^2 +
A(v_\Ls^2 + v_\Rs^2) + Bv_\Ls v_\Rs\right) \ .
\ee
We want $k\neq 0$ at the origin and $k=0$ asymptotically.
The latter is achieved  if
\be
{\partial^2 V\over \partial k^2}\Big|_{k=0}\ =\
m^2 + A \DT^2\ >\ 0 \ .
\label{t_stab}
\ee
At the origin the nontrivial value $k_{(0)}$ becomes
\be
k_{(0)}^2\ {\buildrel x\rightarrow0 \over \longrightarrow}
-{1\over L}\left(m^2+(2A+B)v_{(0)}^2\right)\ ,
\ee
where $v_{(0)}\cong \frac{1}{2}\DT$ is the common value
of $v_\Ls$, $v_\Rs$ at the origin.  Thus
\be
{\partial^2 V\over \partial k^2}\Big\vert_{k=k_{(0)}}\ =\
2Lk_{(0)}^2\ =\ -2\left(m^2+(2A+B)v_{(0)}^2\right)\ >\ 0\ .
\label{nt_stab}
\ee
Comparing eqs.\ (\ref{t_stab}) and (\ref{nt_stab}), both conditions
are satisfied provided the effective coefficient $B$ becomes
sufficiently negative.

We can now proceed to determine a sufficient condition for
a position-dependent nontrivial solution. We have already
restricted ourselves to the case $|k_1|=k_2$. We assume 
that the fates of the separate parts Im($k_1$) and Re($k_1$)
are the same, {\it i.e.}, if one of them is nontrivial, both
would be so. So we focus on the condition for $k$ to be
nontrivial.
If the nontrivial solution is energetically favorable, the
trivial solution should be unstable. Thus consider the
linearized equation for the fluctuation $\delta k$ about the
solution $k=0$. The desired time dependence of the solution is
\be
\delta k\  \sim\  e^{\epsilon_k t}\times \left({\rm spatial\ part}
\right)
\ee
with real parameter $\epsilon_k>0$ for instability of the fluctuation.
Then
\be
- \delta k'' + (m^2 + A(v_\Ls^2 + v_\Rs^2)
+ Bv_\Ls v_\Rs)\, \delta k\ =\ -\epsilon_k^2 \,\delta k
\ee
We compare this with the Schr\"odinger
equation for a bound state wavefunction
\be
-\psi''(x) + V(x)\psi(x) = E\psi(x)
\label{schro}
\ee
Our $V(x)$ has the form of an attractive potential; it approaches
a positive constant value $V_0$ asymptotically, and
$V_0<0$ near the origin due to eq.\ (\ref{nt_stab}).
In the Schr\"odinger equation above, for a bound state, $E<0$
if $V(x)\rightarrow 0$ asymptotically.
In the present case, due to the positive constant value
of $V(x)$ asymptotically, bound states may exist even for $E>0$.
However our stability analysis demands $E<0$, since
we want $\epsilon_k$ to be real. If we ensure that
the $E\sim 0$ solution has at least one node then there will
be a lower energy solution with no nodes, 
the required unstable fluctuation.
In the WKB approximation this condition amounts to
\be
\int_a^b \sqrt{-V(x)}\,dx\ \ge\ {3\pi \over 2}
\label{wkb_stability}
\ee
where $a$ and $b$ are the zeros of $V(x)$.  
Eqs.\ (\ref{t_stab}), (\ref{nt_stab}) and
(\ref{wkb_stability}) constitute one set of sufficient conditions
on the parameter space for a CP violating condensate to occur
within the width  of the domain wall.  They provide the range
to be explored if a full numerical solution were to be attempted.

\section{Effective hamiltonian}\label{sec:toymodel}

In this section we numerically study an effective Hamiltonian
for the likelihood of generating a CP violating condensate.
In a suggestive notation we choose three fields $L$, $R$ and $K$
representing the VEVs of $\dl$, $\dr$ and the electroweak Higgs 
respectively.
The energy per unit area of the wall configuration can be
taken to be

\begin{eqnarray}
H\:  & = & \int dx\: \{\frac{1}{2}|L'|^{2}
+\frac{1}{4}\rho _{1}|L|^{2}(|L|-M)^{2}+\frac{1}{2}(R')^{2}
+\frac{1}{4}\rho _{1}R^{2}(R-M)^{2}+\rho _{3}|L|^{2}R^{2}\\ \nonumber
 &  & +\frac{1}{2}|K'|^{2}+\frac{1}{4}\lambda (|K|^{2}+m^{2})^{2}
+\alpha _{1}|K|^{2}(|L|^{2}+R^{2})+\beta _{1}|K|^{2}({\rm Re}L)R\\ \nonumber
 &  & +\beta _{2}|K|{\rm Re}(KL)R+\beta _{3}{\rm Re}(K^{2}L)R\}
\end{eqnarray} 
where \( L \) and \( K \) are complex and \( R \) is real. 
\(M\) represents the Left-Right breaking mass scale and \(m\) the 
electroweak breaking
mass scale, both including the appropriate finite temperature corrections. 
Thus \(m^2\) is positive. Likewise the other parameters are determined
 by the parameters of the original lagrangian. 
The equations  we get after rescaling the fields by \( M \) are


\begin{eqnarray*}
\frac{d^{2}L_{1}}{dx^{2}} & = &\rho _{1}L_{1}^{3}\left( 1-\frac{1}{2\sqrt{L_{1}^{2}+L_{2}^{2}}}\right)  \\
 &  & +L_{1}\left[ \rho _{1}\left( \frac{1}{2}+L_{2}^{2}-\sqrt{L_{1}^{2}+L_{2}^{2}}-\frac{L_{2}^{2}}{2\sqrt{L_{1}^{2}+L_{2}^{2}}}\right) +2\rho _{3}R^{2}+2\alpha \left( K_{1}^{2}+K_{2}^{2}\right) \right] \\
 &  & +R\left[ \beta _{1}\left( K_{1}^{2}+K_{2}^{2}\right) +\beta _{2}K_{1}\sqrt{K_{1}^{2}+K_{2}^{2}}+\beta _{3}\left( K_{1}^{2}-K_{2}^{2}\right) \right] \\
 &  & \\
\frac{d^{2}L_{2}}{dx^{2}} & = & \rho _{1}L_{2}^{3}\left( 1-\frac{1}{2\sqrt{L_{1}^{2}+L_{2}^{2}}}\right) \\
 &  & +L_{2}\left[ \rho _{1}\left( \frac{1}{2}+L_{1}^{2}-\sqrt{L_{1}^{2}+L_{2}^{2}}-\frac{L_{1}^{2}}{\sqrt{L_{1}^{2}+L_{2}^{2}}}\right) +2\rho _{3}R^{2}+2\alpha \left( K_{1}^{2}+K_{2}^{2}\right) \right] \\
 &  & -R\left[ \beta _{2}K_{2}\sqrt{K_{1}^{2}+K_{2}^{2}}+2\beta _{3}K_{1}K_{2}\right] \\
 &  & \\
\frac{d^{2}R}{dx^{2}} & = & R\left[ \rho _{1}R\left( R-\frac{3}{2}\right) +\frac{\rho _{1}}{2}+2\rho _{3}\left( L_{1}^{2}+L_{2}^{2}\right) +2\alpha \left( K_{1}^{2}+K_{2}^{2}\right) \right] \\
 &  & +\beta _{1}L_{1}\left( K_{1}^{2}+K_{2}^{2}\right) +\beta _{2}\sqrt{K_{1}^{2}+K_{2}^{2}}\left( K_{1}L_{1}-K_{2}L_{2}\right) \\
 &  & \\
\frac{d^{2}K_{1}}{dx^{2}} & = & K_{1}\left[ \lambda \left(
K_{1}^{2}+K_{2}^{2}+\kbd\right) +2\alpha \left( L_{1}^{2}+L_{2}^{2}+R^{2}\right) +2\left( \beta _{1}+\beta _{3}\right) RL_{1}\right] \\
 &  & + K_{1}\frac{\beta _{2}\left( K_{1}L_{1}-K_{2}L_{2}\right) }{\sqrt{K_{1}^{2}+K_{2}^{2}}}
+\beta _{2}RL_{1}\sqrt{K_{1}^{2}+K_{2}^{2}}-2\beta _{3}RL_{2}K_{2}\\
 &  & \\
\frac{d^{2}K_{2}}{dx^{2}} & = & K_{2}\left[ \lambda \left(
K_{1}^{2}+K_{2}^{2}+\kbd\right) +2\alpha \left( L_{1}^{2}+L_{2}^{2}+R^{2}\right) +2\left( \beta _{1}-\beta _{3}\right) L_{1}R \right] \\
 &  & + K_{2}\frac{\beta _{2}\left( K_{1}L_{1}-K_{2}L_{2}\right) }{\sqrt{K_{1}^{2}+K_{2}^{2}}} -\beta _{2}RL_{2}\sqrt{K_{1}^{2}+K_{2}^{2}}-2\beta _{3}RL_{2}K_{1}\\
 &  &
\end{eqnarray*}

\begin{figure}

{\par\centering \resizebox*{0.9\textwidth}{!}
{\rotatebox{0}{\includegraphics{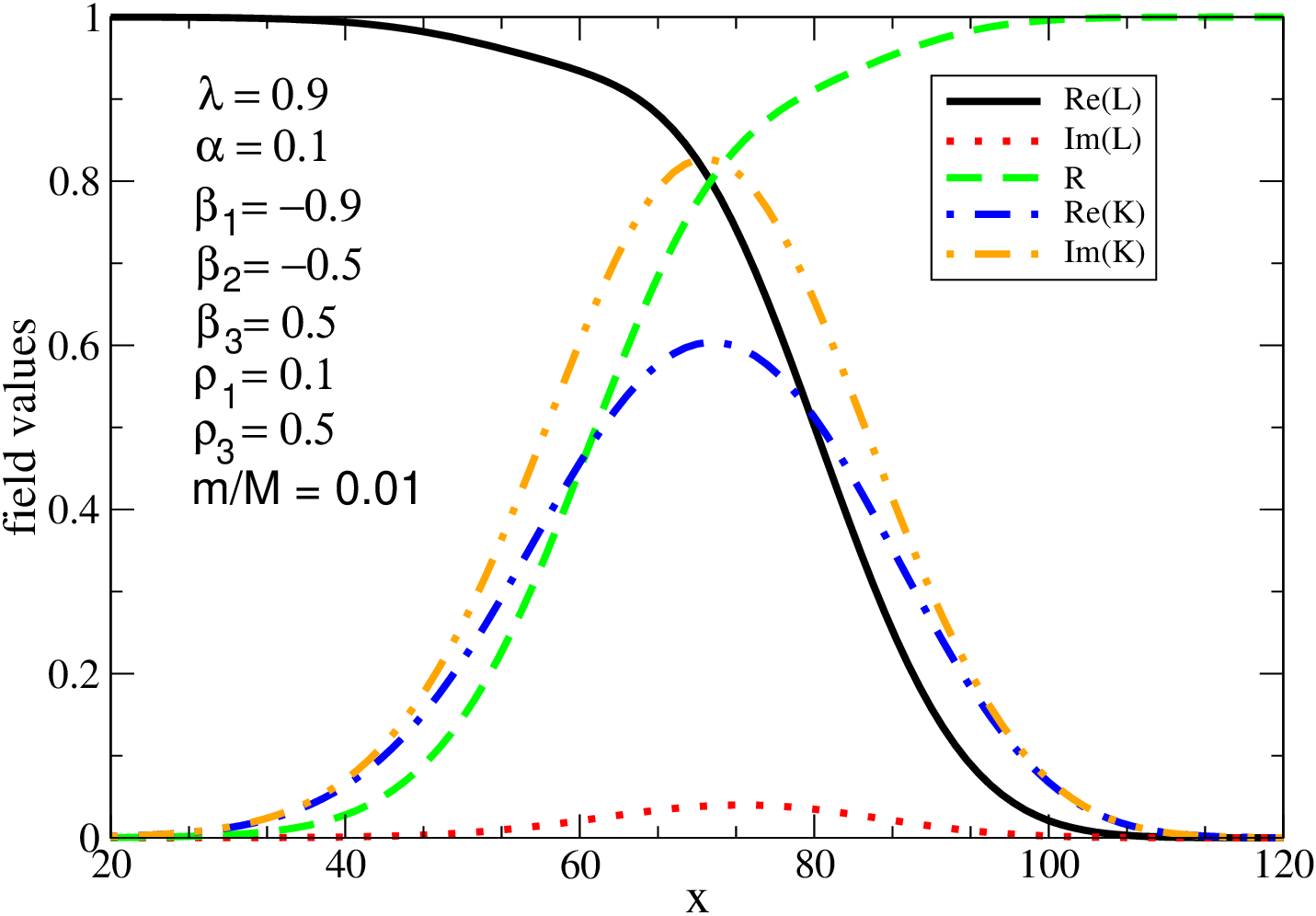}}} \par}\vspace{0.3cm} 

\caption{\(L\), \(R\)  and \(K\) values for given
choice of $\rho_3$, $\alpha_1$ and $\beta_1$, in \(M\) units, x values in 
\(M^{-1}\) units.  Values of parameters not shown are unity.}
\label{fig:fig1}
\end{figure}

\begin{figure}

{\par\centering \resizebox*{0.9\textwidth}{!}
{\rotatebox{0}{\includegraphics{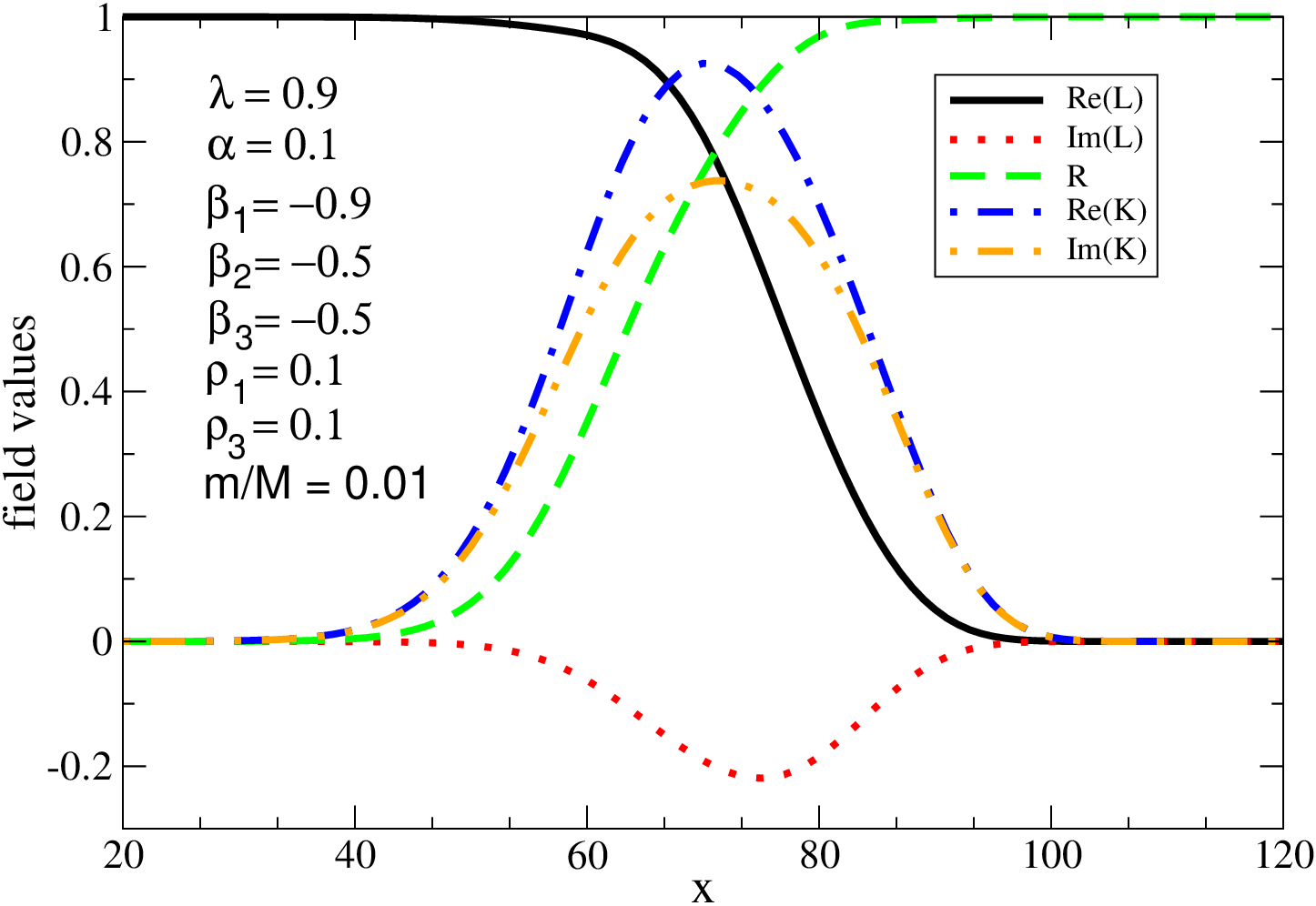}}} \par}\vspace{0.3cm} 

\caption{Same as in fig.\ \ref{fig:fig1}, but for a different choice of
$\beta_1$.}
\label{fig:fig2}
\end{figure}

In addition to the above we need the expression 
\begin{eqnarray}
\label{secder}
{\partial ^{2}V\over \partial |K|^{2}}&=&2\alpha _{1}(|L|^{2}+R^{2})
+2\beta _{1}RL_{1}+2\beta_{2}R\left(L_{1}\cos\alpha-L_{2}\sin\alpha\right)\\ 
\nonumber
&&+ 2\beta_{3}R\left(L_{1}\cos2\alpha-L_{2}\sin2\alpha\right)
+\lambda (3K^{2}+m^2)
\end{eqnarray}
for studying stability issues. This shows that to ensure K=0 asymptotically 
(no E-W breaking at L-R scale) we need 
\[2\alpha _{1}+\lambda \kbd>0.\]
To obtain \( K\neq 0 \) in the core of the wall, we 
again use (\ref{secder}) with the values of \(R\) and \(L_1\) in the core 
estimated to be \(0.5\). This suggests the requirement 
\[\lambda K^{2}=-\lambda \kbd-\frac{1}{2} 
\left(2\alpha _{1}+\beta _{1}+\beta_{2}+\beta_{3}\right)>0\]
This has to be revised in view of the actual values of \(R\) and \(L_1\)
due to backreaction of the \(K\) fields, but serves as a good indicator
to the required range of values.

The second equation clearly suggests taking   \( \beta  \)'s negative. 
In particular \( \beta _{1} \) can be \( {\cal O}(1) \) and negative,
which will ensure the required instability of the \( K=0 \) vacuum  inside
the wall core. Two examples of numerical solutions are shown in figures
\ref{fig:fig1} and \ref{fig:fig2}. Parameters other  than those
displayed are taken to be unity. The shape of the  \(K\) profiles is of
the form of the sech function, as expected for lowest linear 
perturbation in a $\tanh$ background. The
numerical study indicates the field profiles are sensitive to the
parameters governing the Yukawa couplings, but they have no appreciable
variation with respect to the ratio of the mass scales $\kbd$. 

\section{Implications for cosmology}\label{sec:cosmology}
We are now in a position to use the formalism of eqs 
(\ref{eq:force})-(\ref{eq:source}) to estimate the lepton asymmetry generated. 
The asymmetry in local number density is given by
\begin{equation}
\Delta n_{\sss L}\equiv n_\Ls(x) - \bar{n}_\Ls(x) = 
	\frac16\,{\mu_\Ls (x) T^2}
\label{eq:nL}
\end{equation}
where $\mu_\Ls$ satisfies the diffusion equation (\ref{eq:diffeq}).
The general form of the solution to this equation is
\bea
\mu(x) &=& \left\{ \begin{array}{ll} \Bp {e^{\lp x}} + \Bm {e^{\lm x}}
   + B \int_{-\infty}^x dy\, \left[e^{\lp(x-y)} - e^{\lm(x-y)}\right]
	S(y), & x<0 \\ 
\Ap {e^{\rhp x}} + \Am {e^{\rhm x}}
   + \, A \int_{-\infty}^x dy\, \left[e^{\rhp(x-y)} - e^{\rhm(x-y)}
	\,\right]
	S(y), & x>0 \end{array} \right. 
\eea
where
\bea
\lp &=& 0,\quad \lm = -{v_w\over D},\quad \rho_{\sss \pm} = -{v_w\over 2D}
	\pm \sqrt{\left({v_w\over D}\right)^2 + {\Gamma_{\rm hf}\over D}}\\
	B &=& {1\over D(\lm-\lp)} = -{v_w^{-1}},\quad 
	A = {1\over D(\rhm-\rhp)} = 
-\left({v_w^2+4\Gamma_{\rm hf} D}\right)^{-1/2}.
\label{eq:generalform}
\eea
The integration constants $A_{\sss\pm}$ and $B_{\sss\pm}$ are chosen
so that $\mu(x)$ and its derivative are continuous at $x=0$, and 
$\mu$ is finite as $x\to\pm\infty$.  In particular, we are interested
in the limiting value $\mu_0 = \lim_{x\to-\infty}\mu(x) = \Bp$, since
this is relevant deep within the $L$-like phase and represents the
uniform lepton asymmetry filling the universe long after the wall
has passed by.  It is given by
\be
	\mu_0 = {1\over v_w}\int_{-\infty}^0 dy\, \left(1 +
	{\rhp\over\rhm} e^{v_wy/D}\right)S(y)  - {1\over \rhm D}
	\int_0^\infty e^{-\rhp y} S(y).
\ee

We note that in the limit $v_w\to 0$, the above expression is
finite for a generic source $S(y)$.  But since our source vanishes
when $v_w=0$, we get no lepton asymmetry in that limit, which is
in accord with Sakharov's out-of-equilibrium requirement.  We can
also verify that no lepton asymmetry arises when lepton violating
interactions are turned off.  For us, this means setting $\Gamma_{\rm
hf}=0$, in which case we obtain 
$\mu_0 = v_w^{-1}\int_{-\infty}^\infty S(y) dy$.  The integral vanishes
if the source itself does not violate global lepton number conservation,
so this check also succeeds.  The third necessary ingredient, CP
violation, is contained within the source $S(y)$, since this depends
on the neutrino masses having complex phases which vary within the
domain wall.

Now we proceed to estimate the chemical potential $\mu_0$ which 
quantifies the generated lepton asymmetry.  This requires the thermal
averages  \cite{cjk2}
\bea
\label{v2}
\langle \vec v^2 \rangle &=& {3x_\nu+2\over x_\nu^2+3x_\nu+2}\cong 1;\quad
	x_\nu \equiv {m_\nu^2(x)\over T^2} \\
\label{vE2} \left\langle |v_x|\over E^2\right\rangle
	 &\cong& {a - b x_\nu\over  T^2};
	\quad a \cong 0.24;\quad b \cong 0.65
\eea
The first approximation (\ref{v2}) is good for relativistic neutrinos with
$x_\nu\lsim 0.1$, and the second one (\ref{vE2}) is an approximation
to the function given in  \cite{cjk2} which is adequate for
our estimate.  By taking $\langle \vec v^{\,2} \rangle\cong 1$ we can
simplify the expression for $\mu_0$ since the source $S(y)$ becomes
a total derivative.  Integrating by parts,
\be
   \mu_0 \cong {v_w\over T^2}\, {\rhp\over\rhm}\left[\int_{-\infty}^0
   e^{v_wy/D}  + \int_0^\infty e^{-\rhp y}\right] \left(
   \left(a-b {m^2_\nu\over T^2}\right) (m^2_\nu \alpha')' \right) dy
\ee
Since the wall is much thinner than the diffusion scales
$D/v_w$ and $1/\rhp$, it is a good approximation to neglect these
in the integral (setting $e^{v_wy/D}$ and $e^{-\rhp y}$ to 1).  We
will use the ansatz $m^2_\nu(y) = M^2_{\sss N} h^2(y)$, 
$h(y) = \frac12(1+\tanh(y/\Delta_w))$,
for the real part of the neutrino mass, while for the phase,
in accordance with the profiles found in the previous section,
we take  $\alpha(y) = {\rm Im}(L(y)) / {\rm Re}(L(y))$, with
${\rm Im}(L(y))= \alpha_0 \Delta_w h'(y)$.  Here $M_{\sss N}$ 
is the large value of the neutrino Majorana mass neutrino, 
acquired by the left (right) -handed neutrino in the $R$-like 
($L$-like) phase. We have performed
the integral numerically to obtain the analytic result:
\be
	\Delta n_{\sss L} \cong 0.08\, v_w {\alpha_0\over \Delta_w}\, 
	{M_{\sss N}^4\over T^2}
\label{eq:ans1}
\ee
This is the raw value of the lepton number generated
by this mechanism. One would like to 
express this as a ratio $\eta_{\sss L}$ of lepton number to entropy 
as is standard to do with baryon number. Using the expression
for the entropy density ${\cal S} = 2\pi^2 g_* T^3/45$ of $g_*$ 
relativistic degrees of freedom, we get
\be
\label{etaL}
	\eta^{\rm raw}_{\sss L} \cong 0.01\,  v_w {\alpha_0\over g_*}\, 
	{M_{\sss N}^4\over T^5\Delta_w}
\label{eq:ans2}
\ee

Let us consider whether this result can naturally be of the same order
as the observed baryon asymmetry.  
Let $\DT$ [as introduced above,  eq. (\ref{keqn})] denote
the temperature-dependent VEV acquired by the $\Delta_R$ in the phase of
interest.  Experience with the electroweak theory shows that $\DT/T$
is determined by the ratio of gauge and Higgs couplings, and is typically
smaller than unity.  If $f$ is the Yukawa coupling determining
the Majorana mass, then $M_{\sss N} = f \DT$.  Moreover the inverse wall 
width \( \Delta_w^{-1} \) is \( \sqrt{\lambda_{\rm eff}}\, \DT \), 
where \(\lambda_{\rm eff}\) is the effective quartic self-coupling of the 
\(\Delta\) fields.  This is assumed to be small, since we have taken the
wall to be thick.  Therefore we can reexpress (\ref{etaL}) as
\be
\label{etaL2}
 \eta^{\rm raw}_{\sss L} \cong 0.01\,  v_w {\alpha_0\over g_*}\, 
\left({\DT\over T}\right)^5\, f^4\, \sqrt{\lambda_{\rm eff}}\ .
\label{eq:ansparam}
\ee
With $g_*\approx 10^2$, this raw lepton asymmetry is close 
to \( \eta_{\sss B}\cong 10^{-11} \), the desirable value for 
final baryon asymmetry, provided that  
\be
\label{limit0}
\left({\DT\over T}\right)^5\, f^4\, \sqrt{\lambda_{\rm eff}} \approx 10^{-7}\ .
\ee
Even if $\DT/T\sim O(1)$ and $\sqrt{\lambda_{\rm eff}}\sim 1$, this
can be achieved with  a reasonable value for
the Majorana Yukawa coupling \( f_3 \approx 10^{-2} \) 
of the heaviest (third generation) sterile neutrino. 
Ignoring further evolution of the lepton asymmetry for the moment, 
one could turn this around to derive a lower bound on $f$, assuming that
the present mechanism was responsible for baryogenesis.
If all the Majorana
neutrinos are lighter than \( \approx  10^{-2}v_R \), then 
it produces too small a lepton asymmetry to be significant.

After the domain walls have disappeared, the lepton asymmetry 
undergoes further processing by several interactions. Firstly 
the electroweak sphalerons will redistribute this
asymmetry partially into baryonic form. This is the mechanism
by which we get baryon asymmetry from the wall generated lepton
asymmetry. The standard chemical equilibrium calculation 
 \cite{khlebshap} gives
\be
\Delta n_{\sss B}\ =\ \frac{28}{79}\Delta n_{B-L}\ =\ 
-\frac{28}{51}\Delta n_{\sss L}
\label{eq:LtoB}
\ee
assuming the minimal Higgs and flavour content of the Standard 
Model. 

\begin{figure}[bht]
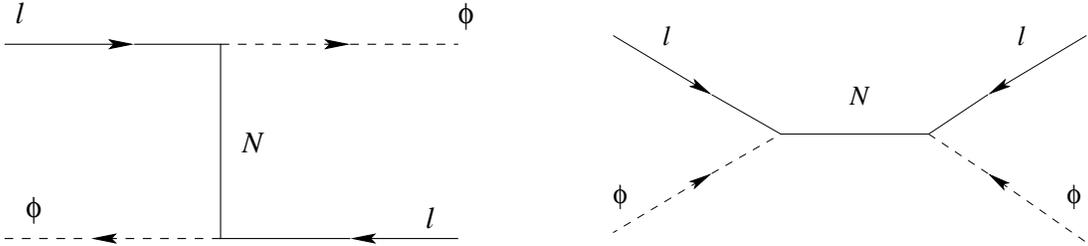

\centerline{\epsfxsize=2.5in\epsfbox{viol1.pstex}\hfil
\epsfxsize=2.5in\epsfbox{viol2.pstex}}
\caption{The $N$ mediated processes violating $L$}
\label{fig:b-l_viol}
\end{figure}

However the presence of heavy majorana neutrinos gives rise
to processes that can deplete the lepton asymmetry generated. Such
processes were considered in a model independent way in 
\cite{luty, Harvey:1990qw, Fischler:1990gn},  
referred to in section \ref{sec:intro}. 
The present model differs
from classic GUT scenarios in that the temperature $T_{B-L} = 
T_{\sss LR}$ 
when the lepton asymmetry is created can be less than or comparable 
to the heavy neutrino mass $M_\Ns$. The two  
processes of importance are the decay of the heavy neutrino
with rate $\GD$ and heavy neutrino-mediated
scattering processes with rate $\GS$. The latter
class of processes in the context of the present model is
shown in fig.\ \ref{fig:b-l_viol}. The rates are given roughly by
\be
\label{rates}
\GD \sim {h^2 M_\Ns^2 \over 16\pi(4T^2 + M_\Ns^2 )^{1/2}}
\hspace{2cm}
\GS \sim {h^4 \over 13\pi^3}{T^3 \over 
(9T^2 + M_\Ns^2)}
\ee
These expressions correctly interpolate between the high and low
temperature limits which can be inferred from eqs.\ (3.1,3.8), (A15-A16)
of ref.\ \cite{luty}, using the Boltzmann approximation 
$K_1(x)\sim e^{-x}/x$ in the thermal
average of the scattering cross section.  (The factor $13$ in (\ref{rates})
is really $96\zeta(3)/9$.)


Let us first consider the case when the decays do not deplete the
generated lepton asymmetry at all. This happens if the lightest of
the heavy majorana neutrinos has $M_{\Ns 1}>T_{\sss LR}$, so
that the decays do not occur because of Boltzmann suppression. This
limit tends to 
make the initial lepton asymmetry $\eta_\Ls$ large, possibly
$O(1)$ from (\ref{eq:ans2}). However the lepton-violating
scattering processes will dilute this by the factor
$
10^{-d_{\sss L}}\equiv\exp\left(-\int_{t_{\sss LR}}^{t_0} \GS dt\right)
$
where $t_{LR}$ is the time of the LR-breaking phase transition and
$t_0$ is the present.  At the same time, sphalerons will keep the baryon
and lepton asymmetries in the same proportion \cite{khlebshap} until 
the electroweak phase transition, at which time the sphalerons go out of
equilibrium.  The corresponding depletion factor for baryons, rewritten
in terms of an integral with respect to temperature, is
\be
	 10^{-d_{\sss B}}\equiv\exp\left(-\int_{T_{\sss EW}}^{T_{\rm min}} {\GS\over H}\,{dT\over T}
	\right);\qquad T_{\rm min}\equiv {\rm min}(T_{\sss LR},\ T_{\rm sph})
\ee
where $T_{\rm sph}\sim 10^{12}$ GeV is the maximum temperature below which
sphalerons are in equilibrium.
Evaluating the integral gives the baryon depletion exponent
\be
\label{dsB}
	d_{\sss B} \cong {3\sqrt{10}\over 13\pi^4\ln10\sqrt{g_*}}\, 
h^4{M_{Pl} T_{\rm min}\,
 \over M_\Ns^2}
\ee
where $g_*$ is the average number of relativistic degrees of freedom,
and we are assuming that $M_{\Ns 1}>T_{\rm min}$.   Eq.\ (\ref{dsB})
can be solved for the Yukawa coupling $h$ which gives the Dirac mass
term for the neutrino: $h^4 \lsim 3200\, d_{\sss B}\left({M_\Ns^2\over
 T_{\rm min} M_{Pl}}\right)$ 
where we have taken $g_* = 110$ for definiteness.  
Since $d_{\sss B}$ should be no greater than about 10 to avoid too much
dilution of the baryon asymmetry, this can be further transformed into an
upper limit on the light neutrino masses using the seesaw relation $m_\nu = 
(hv)^2/M_\Ns$ where $v$ is the Higgs boson VEV, $v = 174$ GeV:
\be
	m_\nu \lsim {180 v^2\over \sqrt{T_{\rm min} M_{Pl}}}\,\left({d_{\sss
B}\over 10}\right)^{1/2}
\ee
If the heaviest neutrino mass is $1$ eV, for example, the temperature of the
$LR$ phase transition (if it is smaller than $T_{\rm sph}$), being
also the temperature at which most of the $B-L$ is generated,
is predicted to be 
\be
\label{limit1}
	T_{B-L}\ =\ T_{\sss LR}\  \lsim 10^{13} {\rm\ GeV} \times \left({{\rm\ eV}\over m_\nu }\right)^2
	\times \left({d_{\sss B}\over 10}\right)
\ee

The previous discussion of dilution by lepton-violating scattering
assumed the heavy neutrinos $N$ had masses $M_\Ns > T_{\sss LR}$ so
that the decay processes could be neglected.  If we are in the opposite
regime, $M_\Ns < T_{\sss LR}$, the decays and inverse decays of $N$
will dominate over scattering for the epoch of temperatures $T > M_\Ns$.
For lower temperatures, the decay rate is exponentially suppressed by
the Boltzmann factor $e^{-M_\Ns/T}$.  We can roughly estimate the dilution
due to decays as 
\be
d_{\sss B} \ = \ {1\over \ln 10}\int_M^{T_{\rm min}}
(\Gamma_D/H) dT/T 
\ \cong\ 	{3\sqrt{10}\over 96\pi^2\sqrt{g_*}\ln 10}\, h^2 {M_{Pl}\over M_\Ns}
	\ \sim 4\times 10^{-4}\ {m_\nu M_{Pl}\over v^2}
\ee
in the limit that $M_\Ns \ll T_{\rm min}$.
Again requiring that $d_{\sss B}<10$ gives the bound on the heaviest
neutrino mass
\be
\label{limit2}
	m_\nu < 0.3\ {\rm eV}\times \left({d_{\sss B}\over 10}\right)
\ee
It is interesting that this value is compatible with, and not very far from
the value implied by atmospheric neutrino observations.

\section{Conclusion}\label{sec:conclusion}

We have shown that a hitherto unexplored mechanism exists in the Left-Right
symmetric model for generation of the observed baryon asymmetry of the
Universe.  The idea is reminiscent of electroweak baryogenesis, but here
the motion of domain walls with CP-violating reflections of neutrinos
during the LR-breaking phase transition creates a large lepton asymmetry,
which is subsequently reprocessed by sphalerons into the baryon asymmetry. 
Unlike electroweak baryogenesis, there is no suppression by $\alpha_W^4$,
since the sphalerons are in equilibrium and they have sufficient time to
equilibrate the baryon and lepton numbers. Rather, the answer is determined
to a large extent by $f^4$ (see (\ref{eq:ansparam})) because asymmetry
production is determined by the helicity flipping interactions. There
are no natural smallness requirements on this parameter, although through
see-saw formula it is constrained by the observed light neutrino mass.
Furthermore there are no strong
constraints on the CP violating phases since they appear in the
interactions of the heavy right handed neutrino.  

It is possible to generate the observed baryon asymmetry for a range of
parameters of the model.  We have studied a few limiting cases
to demonstrate the intrinsic potential of this scenario for producing
the observed baryon asymmetry. One extreme possibility is that we could 
start with the raw value of the lepton asymmetry (\ref{etaL2}) being of order
$10^{-10}$ by virtue of a small Majorana Yukawa coupling, eq.\
(\ref{limit0}), while the heaviest left-handed neutrino mass 
satisfies the bounds (\ref{limit1}, \ref{limit2})
(evaluated at $d_{\sss B}\sim 1$) which guarantee that there is no 
subsequent dilution of the asymmetry by lepton-violating interactions. The
other limiting case is to initially create an  asymmetry of $O(1)$, by
taking large Majorana Yukawa couplings; the asymmetry is subsequently
diluted to the required level by saturating the bounds
(\ref{limit1}, \ref{limit2}), which make reference to the heaviest
left-handed neutrino mass.  

An interesting application of this mechanism is the possibility  to
generate a large lepton number as suggested in  \cite{Harvey:1981cu} and
considered in the context of new observations in cosmology e.g. in 
\cite{Rowan-Robinson:2000ys}-\cite{LL}, notably the MAP and
PLANCK experiments to measure the cosmic microwave background (CMB) fluctuations.
In the simplest model with just one lepton generation, we cannot create a
large lepton asymmetry without also making the baryon asymmetry too large.
But consider a model with a certain combination of lepton numbers 
conserved, such as $L_e + L_\mu$.  This would be the case if the Majorana
mass matrix had the form $({0\ M\atop M\ 0})$.  Then the leptogenesis
mechanism would create equal and opposite amounts of $L_e$ and $L_\mu$.
Since sphalerons separately conserve $\frac13 B-L_e$ and $\frac13 B-L_\mu$,
the combination $\frac23 B-L_e-L_\mu$ would remain conserved at all times,
so that  the resulting baryon asymmetry would be zero even if $|L_e|$ and
$|L_\mu|$ separately were large.  By adding a very small breaking of the 
$L_e + L_\mu$ symmetry, one could generate the observed baryon asymmetry
simlutaneously with large lepton asymmetries \cite{MR}.  
In addition to its imprint on the CMB,  such an effect could
have other observable consequences as observations relevant to nucleosynthesis
are improved \cite{Shi} -\cite{sarkar}.

\section*{Acknowledgment}
This work was started at the 6th workshop on High Energy
Physics Phenomenology (WHEPP-6), IMSc, Chennai,
India. The work of UAY and SNN is supported by a Department of Science
and Technology research grant.
 
\section*{Appendix A: minimization conditions for wall profiles}

These conditions for finding the wall profiles were used in section
\ref{sec:wallprofiles}
\bea
{\delta V\over \delta v_\Ls} &=&
\frac{1}{2}\{\alpha_1 k_1^2  v_\Ls + \alpha_1 k_2^2  v_\Ls
+ \alpha_3 k_2^2  v_\Ls - 2 \mu_3^2 v_\Ls \nonumber \\
&+& 2 \rho_1 v_\Ls^3  + \rho_3 v_\Ls v_\Rs^2
+ 4 \alpha_2 k_1 k_2 v_\Ls \cos(\alpha)  \nonumber \\
&+&  \beta_1 k_1 k_2 v_\Rs \cos(\alpha - \theta)
+ \beta_2 k_1^2  v_\Rs \cos(2 \alpha - \theta)
+ \beta_3 k_2^2  v_\Rs \cos(\theta)\}\\
{\delta V\over \delta v_\Rs} &=&
\frac{1}{2}\{\alpha_1 k_1^2  v_\Rs + \alpha_1 k_2^2  v_\Rs
+ \alpha_3 k_2^2  v_\Rs \nonumber \\
&-& 2 \mu_3^2 v_\Rs + \rho_3 v_\Ls^2  v_\Rs
+ 2 \rho_1 v_\Rs^3   \nonumber \\
&+& 4 \alpha_2 k_1 k_2 v_\Rs \cos(\alpha)
+ \beta_1 k_1 k_2 v_\Ls \cos(\alpha - \theta) \nonumber \\
&+& \beta_2 k_1^2  v_\Ls \cos(2 \alpha - \theta)
+ \beta_3 k_2^2  v_\Ls \cos(\theta)\}\\
{\delta V\over \delta \theta} &=&
-\frac{1}{2}v_\Ls v_\Rs \{-\beta_1 k_1 k_2 \sin(\alpha
- \theta) - \beta_2 k_1^2  \sin(2 \alpha - \theta)
+ \beta_3 k_2^2  \sin\theta\} \\
{\delta V\over \delta k_2} &=&
\lambda_1(k_1^2  k_2  + k_2^3)   + 2 \lambda_3 k_1^2  k_2\nonumber \\
&-& \mu_1^2 k_2
+ \frac{1}{2}\alpha_1 k_2 v_\Ls^2
+ \frac{1}{2} \alpha_3 k_2 v_\Ls^2 \nonumber \\
&+& \frac{1}{2} \alpha_1 k_2 v_\Rs^2
+\frac{1}{2} \alpha_3 k_2 v_\Rs^2
+ k_1^3  \lambda_4 \cos\alpha \nonumber \\
&+& 3 k_1 k_2^2  \lambda_4 \cos\alpha - 2 k_1 \mu_2^2 \cos\alpha
+ \alpha_2 k_1 v_\Ls^2  \cos\alpha
+ \alpha_2 k_1 v_\Rs^2  \cos\alpha \nonumber \\
&+& 4 k_1^2  k_2 l\alpha_2 \cos2\alpha
+ \frac{1}{2} \beta_1 k_1 v_\Ls v_\Rs \cos(\alpha - \theta)
+ \beta_3 k_2 v_\Ls v_\Rs \cos\theta\\
{\delta V\over \delta k_1} &=&
k_1^3  \lambda_1 + k_1 k_2^2  \lambda_1
+ 2 k_1 k_2^2  \lambda_3 - k_1 \mu_1^2\nonumber \\
&+& \frac{1}{2} \alpha_1 k_1 v_\Ls^2
+\frac{1}{2}\alpha_1 k_1 v_\Rs^2
+ 3 k_1^2  k_2 \lambda_4 \cos(\alpha)\nonumber\\
&+& k_2^3  \lambda_4 \cos(\alpha)
- 2 k_2 \mu_2^2 \cos(\alpha)
+ \alpha_2 k_2 v_\Ls^2  \cos(\alpha) \nonumber\\
&+& \alpha_2 k_2 v_\Rs^2  \cos(\alpha) + 4 k_1 k_2^2  \lambda_2 \cos(2 \alpha)
+\frac{1}{2} \beta_1 k_2 v_\Ls v_\Rs \cos(\alpha - \theta)\nonumber\\
&+& \beta_2 k_1 v_\Ls v_\Rs \cos(2 \alpha - \theta)\\
{\delta V\over \delta \alpha}&=&
-\lambda_4 k_1 k_2 (k_1^2  + k_2^2 ) \sin\alpha
+ 2 k_1 k_2 \mu_2^2 \sin\alpha \nonumber\\
&-& \alpha_2 k_1 k_2 (v_\Ls^2  + v_\Rs^2 ) \sin(\alpha)
- 4 k_1^2  k_2^2  \lambda_2 \sin2\alpha
- \frac{1}{2} \beta_1 k_1 k_2 v_\Ls v_\Rs \sin(\alpha - \theta) \nonumber\\
&-& \beta_2 k_1^2  v_\Ls v_\Rs \sin(2 \alpha - \theta)
\eea

\bigskip
\hrule
\bigskip

\end{document}